\documentclass[12pt,a4paper,reqno]{article}
\usepackage{amsmath}
\usepackage{amssymb}
\usepackage{amsthm}

\renewcommand{\(}{\left(}
\renewcommand{\)}{\right)}

\renewcommand{\epsilon}{\varepsilon}
\renewcommand{\hbar}{\hslash}

\renewcommand{\S}{\mathcal{S}}

\newcommand{\ox}{\otimes}
\newcommand{\<}{\langle}
\renewcommand{\>}{\rangle}

\renewcommand{\Im}{\operatorname{Im}}

\theoremstyle{plain} 

\theoremstyle{definition}

\theoremstyle{remark}

\begin{document}
\title{Why am I me?\\ and why is my world so classical?}
\author{Anthony Sudbery\\[10pt] \small Department of Mathematics,    
University of York,\\[-2pt] \small Heslington, York, England YO1 5DD\\    
\small  Email: as2@york.ac.uk}
\date{19 November 2000}
\maketitle
\begin{abstract}
This is an attempt to apply Nagel's distinction between internal
and external statements to the interpretation of quantum mechanics. I
propose that this distinction resolves the contradiction between unitary
evolution and the projection postulate. I also propose a more empirically
realistic version of the projection postulate. The result is a version of
Everett's relative-state interpretation, including a proposal for how
probabilities are to be understood.

Based on a talk given at the 9th UK Foundations of Physics
meeting in Birmingham on 12 September 2000.

\end{abstract}

In this talk I want to explore the possibility of a connection between
the problem of understanding quantum mechanics and a number of classical
philosophical problems. For this purpose, I will at first
focus on the measurement problem, which can be given a conveniently
concise formulation: the contradiction between the Schr\"odinger
equation and the projection postulate. We appear (at least immediately
after reading our first quantum mechanics textbook) to have good reason to
believe both of these incompatible statements. The problems to which I want 
to compare the measurement problem can also be formulated as
contradictions between pairs of statements or principles, both of which
we seem to have good reason to believe:
\begin{enumerate}
\item The existence of space-time \emph{vs} the passage of time;
\item Determinism \emph{vs} free will;
\item The physical description of brain states \emph{vs} conscious experience;
\item Duty \emph{vs} ``Why should I?"
\end{enumerate}
One thing that all of these oppositions have in common with that between
the Schr\"odinger equation and the projection postulate is that in every
case one of the statements is a general universal statement --- what
Nagel \cite{Nagel:nowhere} calls ``a view from nowhere" --- to which
assent seems to be compelled by scientific investigation or moral
reflection; the other is a matter of immediate experience (a view from
``now here"). In other ways
there may seem to be less similarity between the problem of quantum
mechanics and the others. One difference for me personally is that
although the contradiction between the Schr\"odinger equation and the
projection postulate is sharp and uncomfortable, I do not see any
contradiction in most of the other cases. This, of course, is favourable
for my project: if I can succeed in establishing a relation between
all these pairs, then I can hope that the solvent that removes 
the other contradictions will also work on the quantum-mechanical one.
First, however, I have to recall the other contradictions to myself 
so that I can see how they go away. 

For some of my audience, I think, all of the above contradictions have
sharp teeth. In an attempt to bring us all together to share the
experience of summoning up a vanished tension and watching it relax, I
would like to start by considering the problem of my title. This is
slightly different from the others, in that it is not a contradiction,
but it brings us very directly to what I take to be the insight of Nagel
\cite{Nagel:subjobj, Nagel:nowhere}, which he applied to the other contradictions and 
which I will attempt to apply to the interpretation of quantum
mechanics.

\medskip
\begin{center} WHY AM I ME? \end{center}

At the age when philosophy is a natural and urgent activity, children often ask
``Why am I me?". What can they mean by this question? ``I" and ``me" are
different grammatical forms of the same
substantive; they have the same referent. How can it be problematical
that they are identical? Yet what the question seems to be expressing is
the sense that it is \emph{contingent} that I am me. Can that make
sense? Might I not have been Tony Sudbery --- might I have been Mick
Jagger? Clearly not; ``I", when spoken by me, denotes Tony Sudbery, and
Mick Jagger is a different person. It is not logically possible that
these distinct individuals could be equal. But the child in me would say
``Yes, I might have been Mick Jagger" and I understand him; there seems
to be a
sense in which that is true. If so, the subject of the sentence cannot
after all be ``Tony Sudbery". Attempts to identify the true subject of
the sentence might lead me astray, towards ``I who am inside Tony
Sudbery" or ``I who experience the world as Tony Sudbery", but Nagel has
shown a better way. Let us just say that ``I" is not always a simple 
synonym for the (objectively definable) speaker; it sometimes refers to
the experiencing subject. In the world of each of us, there are many 
human beings, but there is only one experiencing subject. It would be a
neat theory to declare that children ask ``Why am I me?" at an age when
they have only just realised that other people also experience the world
as subjects, but I doubt if this is true. The human faculty of
projection, of seeing other human bodies as persons, probably develops before
the faculty of language. Nevertheless, the question
expresses the tension between the knowledge that there are many persons,
each of whom experiences the world in the same way that I do, and the
more immediately known fact that there is only one such experience to which
I can directly attest. It is contingent that that directly attested
experience is what it happens to be in my life; that I (the experiencing
subject) am me (Tony Sudbery).

\medskip
\begin{center} THE FLOW OF TIME \end{center}

Many physicists find a contradiction between their experience of time
and the description of time that they give as physicists. For example:

\begin{enumerate} \item \label{Einstein} For us convinced physicists the
distinction between past, present and future is an illusion, although a
persistent one. (Einstein \cite{Einstein:time})

\item \label{Davies1} Things don't \emph{happen} in spacetime, they
simply \emph{are}. (Paul Davies \cite{Davies:otherworlds})

\item \label{Carr} Relativity \ldots seems incapable of describing the
flow of time at all: past, present and future co-exist in a
four-dimensional ``block", dubbed space-time. (Bernard Carr \cite{Carr})

\item \label{Davies2} There seems to be no strong reason for supposing
that the flow of time is any more than an illusion produced by brain
processes similar to the perception of rotation during dizziness. (Paul
Davies \cite{Davies:otherworlds})

\end{enumerate}

I don't myself share this sense of contradiction. If the flow of time is
an illusion, what is it that we mistakenly believe when we are under
this illusion?\footnote{A similar question can be asked about the
supposed illusion of free will.} Not that time flows, because that
doesn't make sense: only substances flow, and time is not a substance.
Paul Davies's illusion, in fact, is itself illusory. 

Einstein's assertion that there is no distinction between past, present
and future, on the other hand, seems to be simply mistaken.
The past, the present and the future are the sets
of events with time coordinates $t$ satisfying $t<t_0$, $t=t_0$ and
$t>t_0$ respectively, where $t_0$ is a certain time which I can specify
--- 11.20 on 8 September, 2000, actually. It is no illusion that there
is a distinction between these sets; they are distinct sets. What lies
behind this idea of an illusion, of course, is Einstein's discovery that
the distinction differs from observer to observer. This does not alter
the fact that in every frame of reference (and for every choice of
$t_0$) there is such a distinction, and it does not make the distinction
any less real and objective. 

However, arguments about the relativistic meaning of past, present and
future seem to miss the essential point. The immediate application of
these terms is to events in one individual's experience, that is to
events on a particular worldline. The division of these events into
past, present and future is relativistically invariant --- though it
does, of course, depend on the specification of a particular event on
the worldline as ``now".

The other quotations above are even easier to demolish. It is true that things (Davies
means ``events") \emph{are} in space-time, but why does that mean that they
don't \emph{happen}? Happening, as a matter of linguistic fact, is just 
what events do. Finally, if Davies's statement (\ref{Davies2}) is a
meaningless expression of subjective experience, Carr's description
(\ref{Carr}) of objective space-time is clearly self-contradictory, or
becomes so if we replace ``co-exist" by ``exist at the same time", which
is what Carr seems to mean. It is obviously not true that all the events
in space-time exist at the same time, for events are defined by coordinates
$(t,x,y,z)$, and they do not all have the same time $t$.

But having fun with the daft things that physicists say about time
doesn't shed much light on the question of why there is such a
widespread feeling that physicists' notion of space-time contradicts
what we know from experience about time. My comment on Carr's statement
(\ref{Carr}) reduces this contradiction to a confusion between the time
at which an event happens and the time at which it is discussed; but
recognising this distinction doesn't seem to remove the puzzlement which
we often feel when thinking about time in our own lives. It was this
which was expressed by St. Augustine:

\begin{quotation} 
How can the past and future be when the past no longer is and
the future is not yet? As for the present, if it were always present and
never moved on to become the past, it would not be time but eternity.
\end{quotation}

I want to analyse this puzzlement in two different but related ways.
First, there is a sense that statements referring to the present and the
future cannot both be true, because they contradict each other. Consider
a prisoner who is due to be released this weekend. His being in prison
and his being free cannot both be true; hence if it is true that he is
in prison, the future (in which he is free) cannot exist. Logically, the
resolution is easy: the two states have been incompletely described, and
if we complete them by specifying their time they are not contradictory:
being in prison on 8 September is certainly compatible with being free
on 15 September. But in terms of actual experience a sense of
contradiction remains. Our experience is completely specified without a
time label --- we only experience \emph{one thing at a time} --- and
adding time labels does not make imprisonment and freedom compatible
states in our experience. The prisoner, desperately longing for next
week, finds it hard to really believe that he \emph{will be} free
because he all too clearly \emph{is} in prison. The purely intellectual
acknowledgement of the truth of a statement referring to a remote time
is shadowy and pale in comparison to the vivid knowledge of what we are
experiencing \emph{now}. The first is an external statement; the second
is internal.

The second point to note starts as a linguistic one. Augustine points
out that it cannot be true to say that the future exists because it
\emph{is not yet}. This contradicts the physicist's assertion that all
the events of space-time simply \emph{are}. Augustine would say that
this is wrong; it is not true that future events \emph{are}, only that
they \emph{will be}. We cannot make a statement in ordinary language
without giving it a tense. In the mathematical language in which,
fortunately, statements of physics can be expressed, this restriction
does not operate. In this language we can express the prisoner's present
confinement and his future freedom by considering his time-dependent
state $|\psi(t)\>$ and using the tenseless $=$ sign to write
\begin{align*}
&|\psi(\text{8 September})\> = |\text{in prison}\>,\\
&|\psi(\text{15 September})\> = |\text{free}\>.
\end{align*}
Philosophers sometimes enviously adopt this feature of mathematical expression
by inventing a ``tenseless" form of verbs in ordinary language. This
distinction between tensed and tenseless statements can help to explain
the perceived tension between the existence of space-time and our
experience of time: physical statements about events in space-time
(external statements) are tenseless, whereas statements that we make in
space-time are always tensed, with an implied ``now". The reason
why space-time is taken to be an odd idea, contradicting our intuition,
is that statements about it (like Davies's (\ref{Davies1})) are
tenseless statements expressed in a language that has no tenseless
forms. They are therefore falsely understood as tensed statements, as which they
appear commensurate with the tensed statements of our experience, and
may indeed contradict them. My argument that
this contradiction can be resolved depends on the existence of a translation
from a tensed statement, together with its context, to a tenseless
one, in which one moves the context (the identification of ``now") into
the statement; thus 
\[
\text{``I will be free in one week's time" uttered on 8 September}
\]
translates to 
\[
|\psi(\text{15 September})\> = |\text{free}\>.
\]

\medskip
\begin{center}
EXTERNAL SMOOTHNESS, INTERNAL COLLAPSE
\end{center}

I now want to propose that a distinction between internal and external
statements like that between tensed and
tenseless ones, or between an experiencing subject (I) and
a physically identified body (me), can be used to resolve the
contradiction between the two conflicting laws of evolution in quantum
mechanics: the discontinuous, probabilistic change in the state vector
following a measurement, given by the projection postulate, and the
continuous, deterministic evolution given by the Schr\"odinger equation.
The idea, roughly speaking, is that the Schr\"odinger equation is an
external statement, while the projection postulate is an internal one.
More precisely, let $|\Psi(t)\>$ be a time-labelled sequence of states
satisfying the Schr\"odinger equation. Then it is an external (and
tenseless) statement that the world passes through the sequence of
states $|\Psi (t)\>$. Suppose that at early times the state is a product
$|\phi\>|\psi\>$ where $|\phi\>$ is a state of a measuring
apparatus and a conscious observer, while $|\psi\>$ is the state of
the rest of the world (or simply the system being measured by the
apparatus), and suppose the Hamiltonian includes an instantaneous
measurement made at time $t_0$. Then the solution of the Schr\"odinger
equation with this Hamiltonian and these initial conditions will be of the form
\[
|\Psi(t_0 + \varepsilon)\> = \sum_n c_n|\phi_n\>|\psi_n\>
\]
where the $|\psi_n\>$ are eigenstates of the measured observable, $c_n$
are the coefficients in the expansion of $|\psi\>$ in terms of these
eigenstates, and $|\phi_n\>$ is the state of the apparatus and observer
in which the apparatus registers the result $n$ and the observer is
aware of that result. This is an external statement (about the whole
universe), but it is compatible with the internal statement (a tensed
one, with a ``now" of $t_0 + \epsilon$) that after the measurement the
state $|\psi\>$ has jumped to one of the eigenstates $|\psi_n\>$, the
warrant for which is the experienced fact that the observer's state has
jumped to the corresponding $|\phi_n\>$.

The apparatus-observer state $|\phi_n\>$ can be analysed as
\[
|\phi_n\> = |\text{``The result is $n$"}\>|\alpha_n\>
\]
in which $|\alpha_n\>$ is an apparatus state and a ket symbol containing
a quoted statement represents a state of the observer in which they
believe that statement. Thus the external description of the universe is
the superposition
\[
|\Psi(t_0+\epsilon)\> = 
\sum_n c_n |\text{``The result is $n$"}\>|\alpha_n\>|\psi_n\>.
\]
We see very clearly here that it is correct to call the statements of the result of the
measurement ``internal statements": they occur 
\emph{inside} the external statement, as part of the physical
world. They are configurations of a physical system, namely the brain of
the observer. But they are also propositions. What is their status as propositions:
are they true or false? Each is believed by a brain which has observed the fact it
describes, and that fact belongs to reality. As a
human belief, each statement could not be more true. Yet they cannot all
be true, for they contradict each other. I take this to be
characteristic of internal statements in a physical system; the
belief of such a statement is a physical occurrence, and its truth can
only be assessed in the physical context in which it occurs. In the
present situation, such a context consists of a particular component of
the universal state $|\Psi(t)\>$. 

\medskip
\begin{center}
PROBABILITIES
\end{center}

Clearly what I am proposing here is an understanding of Everett's relative-state
interpretation. Notoriously, this interpretation has a problem with the
probability statements of quantum mechanics. The distinction between
internal and external statements opens a new approach to this problem.
There are two aspects to this. First, probabilities are attached to the
results of measurements. But statements about the results of measurements
must be internal --- it is only from a particular perspective that a
measurement \emph{has} a result --- so we must be prepared to accept
that probabilities are only relevant to internal statements. A statement
of probabilities will not be itself an internal statement, but it will
be \emph{about} internal statements.

Secondly, I find that I cannot understand probabilities in quantum
mechanics unless I move to a formulation of the probabilistic law about
the results of experiments which is slightly different from the usual
one: not equivalent to it, but somewhat stronger --- though no stronger,
I believe, than the (unformulated) law which is actually used by the
practitioners of quantum mechanics. I have argued elsewhere
\cite{verdammt} that the conventional postulate, referring to the
results of measurements, is not an adequate description of empirical
reality. Not only is it incompletely specified, relying as it does on an
undefined notion of ``measurement", but it also fails to give any answer
to many experimental questions to which physicists need answers. It
assumes that all experiments can be described as instantaneous
measurements in which the experimenter actively provokes the system
under investigation into providing a result, and consequently changing
its state, by means of an  instantaneous (or at least sharply
time-dependent) intervention. This does not cover, and cannot be adapted
to cover \cite{verdammt}, the common situation in which the experiment
consists of passively observing the system as it spontaneously changes,
and the experimental setup is constant over an extended period of time.

In order to cover this situation of continuous observation, the
probabilistic statements need to be in the form of transition
probabilities. A convincing postulate was proposed by Bell
\cite{Bell:beables}. In a generalised form (\cite{QMPN}, p. 216), it
consists of the assumption that the state of the system is always in one
of a certain set of subspaces $\S_m$, and that it moves stochastically
from one subspace to another with transition probabilities which are
determined by the solution $|\Psi(t)\>$ of the Schr\"odinger equation
as follows. Let $\Pi_m$ be the projection onto the subspace $\S_m(t)$, and
let $|\psi_m(t)\> = \Pi_m|\Psi(t)\>$. Then at time $t$ the system is in
one of the states $|\psi_m(t)\>$, and if it is in $|\psi_m(t)\>$ at time
$t$ then the probability that it will be in $|\psi_n(t+\delta t)\>$
(where $n\neq m$) at time $t+\delta t$ is $T_{nm}\delta t$ where
\begin{align}
T_{mn} &= \frac{\max(J_{mn},0)}{\<\psi_n(t)|\psi_n(t)\>},\\
J_{mn} &= 
\frac{2}{\hbar}\Im\<\psi_m(t)|H|\psi_n(t)\>. 
\end{align}
(This has been generalised to the case of time-dependent $\Pi_m$ by
Bacciagaluppi and Dickson \cite{BacciaDickson}.) It follows from these
transition probabilities that the usual probabilities for the results of
measurements of $\Pi_m(t)$ hold at all times if they hold at any one
time. 

At first sight there is a deeply unattractive and implausible feature of
this proposal: it depends on the choice of the subspaces $\S_m(t)$,
i.e.\ on a choice of preferred observable. Thus it appears to break the general unitary
symmetry of quantum mechanics. But in the framework I am proposing here
this is no vice. A report of an experimental result is an internal statement, made by a
physical system which is capable of formulating
propositions about its environment and having attitudes of belief
towards those propositions --- in short, a conscious system.
(I use the word ``conscious" reluctantly, 
because I do not want to be understood as restricting the discussion to
human beings, but it seems to be what I mean). A general statement about
experimental results must therefore be made relative to the conscious
system which reports the results. A conscious system,
regarded \emph{as} a conscious system, automatically defines a preferred set of
subspaces, namely those consisting of states in which the conscious
system has definite experiences  --- what Lockwood \cite{Lockwood:manyminds}
calls the \emph{consciousness basis}, though I will use the term
\emph{experience basis}.\footnote{In passing, let us note that an answer to the
question ``Why don't we see superpositions of macroscopic states?" is
that there is no experience state describing such seeing. A
superposition of two experience states is not, in general, an experience
state.} If we are making general statements about the
experiences of conscious systems, there is no loss of unitary symmetry
in making each statement depend on the experience basis of the system to
which it refers. This is the same as the way that statements about
energy in special relativity are necessarily relative to a particular
frame of reference; nevertheless, general statements about energy (for
example, the conservation of energy) are possible and do not break
relativistic invariance. 

We thus arrive at the following general formulation of the laws of
motion in quantum mechanics
:

\begin{enumerate}
\item The universe is described by a time-dependent state vector 
$|\Psi(t)\>$ in the universal state space $\S$,
which evolves according to the Schr\"odinger equation. 
\item \label{transprob} The experience of any conscious subsystem $C$ of the universe is 
described at any time $t$ by a state vector $|\phi_n\>$ in the experience basis
of that subsystem's state space $\S_C$. If this experience is described by
$|\phi_n\>$ at time $t$, then the probability that it is described by
$|\phi_m\>$ at time $t+\delta t$ is $T_{mn}\delta t$ where
\begin{align}
T_{mn} &= \frac{\max(J_{mn},0)}{\<\psi_n(t)|\psi_n(t)\>},\\
J_{mn} &= 
2\Im\left[\hbar^{-1}\(\<\phi_m|\<\psi_m(t)|\)H\(|\phi_n\>|\psi_n(t)\>\)\right] 
\end{align}
and the states $|\psi_n(t)\>$ are the states of the rest of the world,
(elements of $\S_R$ where $\S = \S_C\ox\S_R$),
which are the coefficients of the experience basis states in the
expansion of the universal state vector with respect to this basis:
\[
|\Psi(t)\> = \sum_n|\phi_m\>|\psi_n(t)\>.
\]
\end{enumerate}

How does this solve the problem of probabilities in the relative-state
interpretation? I haven't yet defined what I mean by the probabilities
that occur in (\ref{transprob}) above. But I don't know how to define
probabilities in \emph{any} physical theory. Popular definitions of
probabilities are couched in terms of frequencies --- but then they seem
to me to be wrong, or at best circular; or in terms of degree
of belief --- but then they are not appropriate to physics. If it is
to have the kind of objective meaning that is needed in physics,
``probability" has to be taken as a primitive, undefined, term.
Moreover, for its use in physics maybe one has to restrict the sorts of
thing to which the word can be applied. I'm not sure that I know what
would be meant by the objective probability of a proposition being true.
I have the firmest sense that I understand objective probabilities when
they refer to events happening. Then one could possibly essay a
definition of the objective probability of a (future) event as ``the
degree of expectation of the event which is rational for a fully
informed observer", though this is more an elucidation
than a definition. According to the version of quantum
mechanics being proposed here, events happen in the experience of a
conscious system (and only there, in the case of events like quantum
jumps). It is therefore appropriate, and I claim comprehensible, for
probabilities to occur in this part of the theory; and they do so
as primitive terms, but with the meaning that I have indicated above.

\medskip
\begin{center}
SUBJECTIVE AND OBJECTIVE
\end{center}

Except in the preceding paragraph (where my use of them will probably
strike some of my audience as odd; I will explain), I have tried to
avoid the words ``subjective" and ``objective", though it might seem
natural to use them for the distinction between what I have called
``internal" and ``external" statements (indeed Nagel uses them in
the title of one of his early papers on the topic \cite{Nagel:subjobj}).
I think this would be a mistake. When this distinction is applied to quantum
mechanics as I have been trying to do, the class of internal statements
includes all the statements that we are used to making in classical
physics (more precisely: it includes translations of all the external
statements of classical physics, using the type of translation between
internal and external that I discussed in connection with tensed and
tenseless statements). Among such statements there is already a
distinction between subjective and objective, so there are both
subjective internal statements and objective internal statements. For
example, I would make the subjective statement that the hair of most people in
this room is a kind of green, though the objective fact, I
understand, is that the colours in question are various shades of brown.
Both statements are internal statements, made from inside one particular
component of the state vector of the universe.

This is why I used the term ``objective probability", even though
probabilities only refer to internal statements. I want to make the same
distinction between objective probabilities as physical facts and
subjective probabilities as degrees of belief that we would make in a
classical stochastic theory. In ignorance of the way that
Schr\"odinger's diabolical experimenter prepared his device with the cat
--- not knowing which radioactive material he used, or how much of it,
or how lethal was the poison --- I might be prepared to state how
probable I think it is that the cat will survive after an hour in the
box, and to bet in accordance with that subjective probability; but
there is a fact of the matter about how many radioactive nuclei there
are in the box, and what their half-life is, and this determines the
actual objective probability that the cat will survive for an hour. To
spell this out in recognition of the internal nature of my statements, I
should call this the probability that my experience in the next hour
will not include a transition to seeing the cat dead. But since
\emph{all} my statements about the world are internal --- they cannot be
anything else --- I am entitled to say (or at least to assume, and I
will usually be right) that if I see the cat dead, then it \emph{is}
dead. My failure to spell out pedantically the internal status of my
utterances is not just a shorthand but a justified assessment of what
constitutes reality.

\medskip
\begin{center}
REALITY
\end{center}

Which describes reality --- the internal view or the external view? Both
seem to have a good claim. As objective scientists, we might want to say
that the  external perspective is one which describes the whole of
reality, whereas the internal perspective gives a partial or misleading
view. This, we might consider, is the deep and true reality which
quantum mechanics has revealed to us: all the components of the
universal state vector (the ``many worlds", to use a familiar but
unhelpful phrase) really exist, and it is only because of the
limitations of our perceptual apparatus that we are not directly aware
of them. On the other hand, one could take the view that the first
allegiance of a scientist is to the results of experiments; if anything
is real, they are. Thus \emph{one} of the many worlds is real; the
others form a shadowy sort of potential\footnote{``Potential" here is a doubly
appropriate word. The other components of the state vector represent
outcomes which were potential but are not (from the internal
perspective) actual; and they contribute to the evolution of the actual
state in a way which is similar to the contribution of a potential
function to the motion of a particle.} which governs
the evolution of the real world.  

If there is a dispute here, it is surely a barren one. It doesn't
\emph{matter} which of $|\Psi(t)\>$ or $|\phi_n(t)\>$ on page 9 we call
``real". This must mean that the two assertions ``$|\Psi\>$  describes
reality" and ``$|\psi_n(t)\>$ describes reality" are compatible, which
is to say that ``reality" has different senses in the two statements;
and this is understandable if one of the statements is an internal one
and the other is external.

It may be necessary to insist that abandoning the projection postulate
in the external view of the universe is not to deny the reality of the
projected component in which we find ourselves. Nagel also
\cite{Nagel:nowhere} has emphasised that internal statements should not
be regarded as less true or complete than external ones. Indeed, there
are some truths to which there can be no access from an external
perspective, and no expression as external statements; they  are none
the less truths. 
Among these, for example, are facts about the quality of experience;  I
suggest that facts about the
outcome of experiments have a similar status.
We have already noted that the various statements of the outcomes of an
experiment, although they contradict each other when viewed externally,
are all as true as an empirical statement can be when viewed in their
own context. It is no denial or denigration of the reality of an
experimental result to say that there is a description of the world ---
not just a different description, but a different {\em kind} of
description --- in which the result features as just one component of a
superposition of state vectors.

A similar contextualisation of truth and reality has been
advocated by Simon Saunders, whose understanding
of quantum mechanics in terms of decoherent
histories \cite{Saunders:Synthese1, Saunders:Synthese2, 
Saunders:Synthese3} is close to the interpretation proposed here.

\medskip
\begin{center}
HOW MANY WORLDS? HOW MANY MINDS?
\end{center}

Describing the Everett interpretation in terms
of many worlds or many minds only exposes one unnecessarily to charges of
metaphysical profligacy. Orthogonal components of a state vector are not
separate worlds. There is only one world, and it has one state vector.
That state vector contains a number of possible experiences, but no more
minds than one would expect from a classical description of the world
(one, if one is a solipsist; the number of fertilised human ova, if one
is a Catholic; the number of human beings beyond one's favourite stage
of development, if one is a speciesist; ...). From an
external point of view, there are a number of possible experiences; many
of those experiences belong to the same mind. It seems to me to be a
mistake to describe this multiplicity of experiences as a multiplicity of
minds \cite{Lockwood:manyminds, Albert:book, Donald:progress}, since it
makes the correspondence between minds and brains many-to-one. The
mistake is to confuse brain \emph{states} with brains.

This mistake is similar to one made by Deutsch \cite{Deutsch:book}, in
which form it can perhaps most clearly be seen to be a mistake. Deutsch
takes the two-slit experiment with electrons to demonstrate that each
electron going through one of the slits must be knocked off course by
\emph{another} electron which went through the other slit. This electron
lives in ``another world", and is analogous to the other minds
postulated by many-minds theorist. But the state of the universe in
the two-slit experiment, with one electron, is in an eigenstate of
electron number, and the eigenvalue is 1. There is only one electron.
Equally, there is only one brain for each human observer, and therefore
only one mind.

\begin{center}
SUMMARY
\end{center}

\begin{enumerate}

\item One must distinguish between an external statement about a
physical system and internal statements made within the system. An internal
statement is necessarily relative to a particular perspective.

\item A perspective in a quantum system consists of:
\begin{enumerate}
\item a conscious system;
\item an instant of time;
\item an eigenstate of experience.
\end{enumerate}

\item An external statement about a quantum system (in particular, the
universe) consists of a description of its state vector as a function of
time. The physical law to which this statement is subject is that the
state vector obeys the Schr\"odinger equation.

\item An internal statement in a quantum system is a description of an
experience of a particular conscious system at a particular time. The
physical law to which this statement is subject is that the experience
state changes stochastically according to the transition probabilities
\eqref{transprob}.

\end{enumerate}

It is often said that among the problems concerning the relation between
classical and quantum mechanics is the question ``What determines which
one of the classically allowed states is in fact actualised?"
\cite{Clarke}. If we accept that physics has been forced to abandon 
determinism, we should not be
surprised that there is no answer to this question; but there is still a
puzzle about the way in which indeterminism is incorporated by quantum
mechanics, with its apparently deterministic equation of motion. The
argument of this paper has been that the relation between different
classical outcomes to a quantum experiment is analogous to the relation
between different instants of time (an analogy which has also been made
by Lockwood \cite{Lockwood:manyminds}) and also to the relation between
different centres of consciousness. We should no more expect to be able
to answer ``Why did the experiment have that outcome?" than ``Why is it
now {\em now}?" \cite{Lockwood:book} or ``Why am I me?"

\section*{Acknowledgement}

I would like to acknowledge the inspiration I drew from Tom Baldwin's
inaugural lecture, \emph{Inside Out or Outside In?} Professor Baldwin
should not, of course, be held responsible for the amateurishness of the
philosophising that his lecture inspired in an amateur. I am also
grateful to James Ladyman, Harvey Brown and Charles Francis for
helpful comments and for directing me to some relevant literature.


\end{document}